\documentclass[10pt]{iopart}
\usepackage{amsfonts}
\usepackage{amssymb}

\def\a{\alpha}
\def\b{\beta}

\def\d{\delta}
\def\ep{\varepsilon}

\def\vp{\varphi}

\def\a{\alpha}
\def\b{\beta}

\def\d{\delta}

  \def\D{{\mathcal{D}}}

 \eqnobysec
 
\begin{document}

 \title{Low-dimensional electric charges. Covariant description}

\author{Yakov Itin}

\address{Institute of Mathematics, Hebrew University of
   Jerusalem \\
   and Jerusalem College of Technology}
\ead{itin@math.huji.ac.il}






 \begin{abstract}
A  compact and elegant description of the electromagnetic fields in media and in vacuum is attained in the differential forms formalism. This description is explicitly invariant under diffeomorphisms of the spacetime so it is suitable for arbitrary curvilinear coordinates. Moreover, it is independent of the geometry of the underline spacetime. The bulk electric charge and current densities are represented  by twisted non-singular differential 3-forms. The charge and current densities with a support on the low dimensional submanifolds (surfaces, strings and points) naturally require singular differential forms. In this paper, we present a covariant metric-free description of the  surface, string and point densities. It is shown that a covariant description requires Dirac's delta-forms instead of delta-functions. Covariant metric-free conservation laws for the low-dimensional densities are derived.  
\end{abstract}
\date{\today}

\section{Introduction: Maxwell equations for differential forms}
The 3-dimensional description of the electromagnetic field in media is given in term of four  vector fields \cite{Zom}--\cite{Jack}. Two of these fields, the electric field strength $\bf{E}$ and the magnetic  field strength    $\bf{B}$, are of an intensity nature \cite{Zom}. In the 4D-tensorial formalism, these fields are treated as the components of an antisymmetric second order tensor of the electromagnetic field strength $F_{ij}$. Two other vector fields, the electric excitation $\bf{D}$ and the magnetic excitation $\bf{H}$, are both of a quantity nature \cite{Zom} and assembled in a 
tensorial density ${\cal H}_{ij}$. 

In the differential form description, the tensor fields are replaced by the differential forms 
 \begin{equation}\label{fields}
 F=\frac 12 F_{ij}dx^i\wedge dx^j\,, \qquad {\cal H}=\frac 12{\cal H}_{ij}dx^i\wedge dx^j\,.
\end{equation}
Here the electromagnetic field strength $F$ is an untwisted differential form, while the electromagnetic excitation $H$ is a twisted form.  The necessity  to have two different types of differential forms in electromagnetism and the properties of these forms recently discussed in \cite{IOH}. The key point  is the twisted nature of the form electric current form which  is  exhibited in a proper definition of the corresponding integral quantities.  
 
In differential forms notation, the  Maxwell system  is written in a compact   form 
 \begin{equation}\label{Max1}
 dF=0\qquad d{\cal H}=J\,,
\end{equation}
which is explicitly invariant under the diffeomorphism of the manifold. Thus the equations (\ref{Max1}) valid in arbitrary curvlinear coordinates on the flat Minkowski manifold as well as on a curved pseudo-Riemannian manifold of general relativity. 
A most important feature of the field equations (\ref{Max1}) is their completely independence of the metric structure at all, see \cite{Post}, \cite{birkbook} and the references given therein. It is well known that the equations (\ref{Max1}) are not enough for unique determination   of the fields for the given sources. 
The system must be endowed with the constitutive relations between the fields $F$ and $\cal H$. In vacuum electromagnetism, these relations are constructed from the components of the metric tensor. For the material media, the relation determined by the material constants.  The last case justifies the importance of the metric-free description. 

In this paper, our main attention is to the 4D electric current 3-form $J$.  
Due to  (\ref{Max1}) it is  a twisted 3-form, which coordinate representation is given by  
 \begin{equation}\label{curr}
 J=\frac 1{3!} J_{ijk}dx^i\wedge dx^j\wedge dx^k\,.
\end{equation}
The vector notation of the current is reinstated by the definition $J^i=\ep^{ijkl}J_{jkl}$, where the permutation pseudotensor $\ep^{ijkl}$ is used. Now the  components $J^0=\rho, J^\a=j^\a$ represent the electric 3D-scalar charge and the electric 3D-vector current correspondingly.  

For a bulk  electric current, $J$ is defined by  four {\it smooth functions} $J_{ijk}(x)$. Phenomenologically this current is derived by a suitable averaging of the pointwise electric charges and currents of electrons and ions in  media. 

Another types of charges and currents with a support on the lower dimensional submanifolds such as surfaces and strings are also necessary. A well known example is the free charges and currents in conductors which are good described as  being completely localized on the boundary of a sample. Such charges and currents must be described by {\it singular functions} and correspondingly  by {\it singular differential forms.} A mathematically precise formalism of singular tensor distributions was worked out long ago \cite{de Rham}, \cite{Whit} and \cite{Lich}. This formalism, however, uses explicitly the metric structure, the normal vectors,  or the  volume elements structure, see \cite{Dray:1993xu}, \cite{Dray:1995fb}, \cite{Hartley:1997yc}. 
 
In this paper, our aim is to remove the geometric dependence from the definitions of the  electric charge and current  and to represent these quantities in a metric-free diffeomorfism invariant form.   Integral and differential conservation  laws for  surface, string and point-wise electric currents are derived. 

{\it Notations:} The coordinate  indices are denoted by Roman  letters which run over the range of     $i,j,\cdots=0,1,2,3$, the comma denotes the partial derivatives relative to the coordinates $\{x^0,x^1,x^2,x^3\}=\{t,x,y,z\}$. The Greek indices will be used for the spatial coordinates, $\a,\b,\cdots=1,2,3$. The 4-dimensional
Levi-Civita's permutation pseudo-tensor $\ep^{ijkl}$ is normalized
as $\ep^{0123}=1$. 
\section{Surface charge and its conservation}
\subsection{Surface and Dirac's delta-form}
Consider a 2-dimensional surface in the 3-dimensional position space. For instance, 
it can represent a surface between two media or a boundary of a material sample. We want to study the  surface in its arbitrary motion, so we must  necessarily deal  with the 3-dimensional image of the surface 
in  the spacetime. 
A most general description of  such a hypersurface is given by a scalar 
equation
\begin{equation}\label{surf}
\vp(t,x,y,z)=0\,,\quad \mbox{\rm or, briefly,  }\quad \vp(x^i)=0\,.
\end{equation}
We restrict to   smooth hypersurfases so the function $\vp(x^i)$ is assumed to be differentiable  with a  non-zero gradient,  $\vp_{,i}\ne 0$. 

In order to describe the electric charges localized on the hypersurface, 
we need to involve the Dirac delta-function $\d(\vp)$, which is non-zero only for $\vp(x^i)=0$. 
However, for a function $\vp(x^i)$, described a hypersurface, $\d(\vp)$ is not a well defined physical quantity. It can be seen already from the dimensional analysis. 
For an arbitrary positive constant $k$, we have an identity $\d(k\vp)=(1/k)\d(\vp)$. 
It means that $\d(\vp)$ has a physical dimension which is inverse to the dimension 
of $\vp$. But  the dimension of a function $\vp$, appearing in the homogeneous equation (\ref{surf}),
 is not defined. In fact, we can describe the same hypersurface giving 
an arbitrary dimension to  $\vp$. Certainly a term without a fixed physical dimension  cannot represent a physically meaningful quantity.
 
Moreover, for an arbitrary strictly monotonic function $f$ with a nonzero derivative, the equations $\vp=0$  and $f(\vp)=0$ describe the same hypersurfaces. However, the Dirac delta-functions $\d(\vp)$ and $\d(f(\vp)$ are different.

In order to remove a non-controlled physical dimension of  $\d(\vp)$ and its dependence of the representation $f(\vp)$, we define a 1-form $\D(\vp)$
\begin{equation}\label{delta-form}
{\cal D}(\vp)=\d(\vp)d\vp\,,      
\end{equation}
which must be called as {\it Dirac's delta-form}.   
This quantity is clearly dimensionless, independently of the dimension of $\vp$. Moreover, for a strictly increasing function $f$, 
\begin{eqnarray}\label{delta-form-invar}
{\D}(f(\vp))&=&\d(f(\vp))df(\vp)=\frac 1{f'(\vp)}\d(\vp)f'(\vp)d\vp=\d(\vp)d\vp\nonumber\\&=&\D(\vp)\,.     
\end{eqnarray}
Thus delta-form is independent of the  functional representation.  
We conclude that  a physical sense can be  given namely to the 1-form $\D(\vp)$, not to the function $\d(\vp)$. 

Similarly to the known representation of the Dirac delta-function as a derivative of the step-function, 
the Dirac delta-form can be viewed  as an exterior differential of the step-function $u(\vp)$
 \begin{equation}\label{step}
\D(\vp)=d(u(\vp))=\d(\vp)d\vp\,.
\end{equation}
In a coordinate basis, 
\begin{equation}\label{delta-coord}
\D(\vp)=\d(\vp)\vp_{,i}dx^i\,.
\end{equation}
Thus we conclude that in tensorial equations (written with explicit indices) the term $\d(\vp)\vp_{,i}$ is admissible but not $\d(\vp)$ itself. It worth to remember that a precise definition of the delta-function treats it as a measure which appears into an integral with the definite property
\begin{equation}\label{delta-def}
\int_{-\infty }^\infty  f(\tau)\d(\tau)d\tau=f(0)\,.
\end{equation}
Already in this basic equation, the delta-function emerges  together with the differential $d\tau$, so only the 1-form $\D(\tau)=\d(\tau)d\tau$ is meaningful. 
\subsection{Surface current}
We are able now to deal with  the singular 3-form of electric surface current  in the 4-dimensional
spacetime.  For a surface $\vp(x)=0$, we define a twisted form 
\begin{equation}\label{delta-coord-1}
^{(sur)}\!J=L\wedge\D(\vp)=L\d(\vp)\wedge d\vp\,,
\end{equation}
where $L$ is an arbitrary twisted regular 2-form.  This representation explicitly guarantees the principle properties of the surface current:
\begin{itemize}
\item[(i)] The current must be localized only on the hypersurface $\vp=0$. It is provided by the factor $\d(\vp)$, so for $\vp\ne0$ the current vanishes.

\item[(ii)]  The current must be tangential to the hypersurface. This property is guaranteed by the factor $d(\vp)$. In fact, we have $^{(sur)}\!J\wedge d\vp=0$. 

\item[(iii)] The absolute dimension  of the 2-form $L$ is the same as of the 3-form $^{(sur)}\!J$, i.e., the dimension of charge. For a detailed  discussion of the natural dimensions of the physical quantities described be differential forms, see \cite{Hehl:2004jn}.
\end{itemize}

To have the coordinate expression of the surface current we recall   the usual representation of the 2-forms
\begin{equation}\label{L-coord}
L=\frac 12 L_{ij}dx^i\wedge dx^j\,.
\end{equation}
When it is combined   with (\ref{delta-coord}), we have 
\begin{equation}\label{J-coord}
^{(sur)}\!J=\frac 12 L_{ij}\vp_{,k}\d(\vp)dx^i\wedge dx^j\wedge dx^k\,.
\end{equation}
Comparing it with the ordinary representation of the 3-forms (\ref{curr}) we obtain
\begin{equation}\label{J-coord1}
^{(sur)}\!J_{ijk}= L_{[ij}\vp_{,k]}\d(\vp)\,,
\end{equation}
where the antisimmetrization of the indices is denoted by the square parenthesis. 
The corresponded dual tensor $J^i=\ep^{ijkl}J_{jkl}$  takes the form
\begin{equation}\label{J-coord2}
^{(sur)}\!J^i= L^{ij}\vp_{,j}\d(\vp)\,,
\end{equation}
where the dual $L^{ij}=(1/2)\ep^{ijkl}L_{kl}$ is involved. 

A proper physical sense of the surface current is obtained when the current 3-form is treated as an integrand of an integral defined on a 3-dimensional surface, which we denote by $M_3$. 
The total charge contained in $M_3$ is given by 
\begin{equation}\label{tot-charge1}
Q=\int_{M_3}{}^{(sur)}\!J^i\,.
\end{equation}
Substituting here (\ref{delta-coord}) we have 
\begin{equation}\label{tot-charge2}
Q=\int_{M_3}L\wedge \D(\vp)=\int_{M_3}L\wedge \d(\vp)d(\vp)\,.
\end{equation}
Observe that, if the intersection 
\begin{equation}\label{intersect}
M_2={M_3}\cap\{\vp=0\}
\end{equation}
is empty, the integral is zero. Moreover, due to the usual property of the $\d$-function, the total charge is expressed as an integral of a regular 2-form $L$ over $M_2$ 
\begin{equation}\label{tot-charge3}
Q=\int_{M_2}L\,.
\end{equation}
This integral of a 2-form is non-zero only if the set $M_2$ (the intersection of two 3-dimensional submanifolds) is a 2-dimensional submanifold. Observe that both sides of (\ref{tot-charge3}) the physical dimension of  charge. We will see in the sequel how (\ref{tot-charge3}) is expressed in a natural coordinate system. 
\subsection{Surface charge conservation}
For a surface current defined above, the conservation law is straightforwardly derived from the ordinary charge conservation equation 
\begin{equation}\label{J-cons}
d\left({}^{(sur)}\!J\right)=0\,. 
\end{equation}
Use (\ref{delta-coord}) to calculate
\begin{equation}\label{J-cons1}
d\left({}^{(sur)}\!J\right)=dL\wedge \D(\vp)+L\wedge d\D(\vp)\,.
\end{equation}
Since the $\d$-form is the differential of the step-function, its exterior derivative is equal to zero. Consequently, we have the surface charge conservation law in the form 
\begin{equation}\label{J-cons2}
dL\wedge \D(\vp)=0\,.
\end{equation}
In a coordinate chart, it takes the form
\begin{equation}\label{J-cons-coord}
L_{[ij,m}\vp_{,k]} \d(\vp)=0\,,
\end{equation}
or in term of the dual tensor
\begin{equation}\label{J-cons-coord1}
L^{ij}{}_{,i}\vp_{,j} \d(\vp)=0\,.
\end{equation}
The integral representation of this law is obtained by Stokes theorem. For a 4-dimensional spacetime region 
$M_4$ with a boundary $\partial M_4$, we write it as 
\begin{equation}\label{int-cons1}
\int_{M_4}d\left({}^{(sur)}\!J\right)=\int_{\partial M_4}{}^{(sur)}\!J\,.
\end{equation}
Thus, the conservation law takes the form
\begin{equation}\label{int-cons2}
\int_{\partial M_4}{}^{(sur)}\!J=\int_{\partial M_4}L\wedge\D(\vp)=0\,.
\end{equation}
Providing the integration with respect to the $\vp$ factor we remain with an integral conservation law
of the form 
\begin{equation}\label{int-cons3}
\int_{ M_2}L=0\,,
\end{equation}
where the two dimensional surface is defined as an intersection
\begin{equation}\label{int-cons4}
{ M_2}={\partial M_4}\cap\{\vp=0\}\,.
\end{equation}
 
The different forms of the charge conservation law given above are related to the case when the charges are completely localized on the surface and the bulk charge is zero. Sometimes it is useful to consider a model which inqludes both types of the phenomenological charge, the surface charge density     ${}^{(sur)}\!J$ and the bulk charge ${}^{(bulk)}\!J$. In this case, (\ref{J-cons}) is modified as \begin{equation}\label{J-cons-x}
d\left({}^{(sur)}\!J+{}^{(bulk)}\!J\right)=0\,. 
\end{equation}
Consequently, (\ref{J-cons2}) takes the form
\begin{equation}\label{J-cons2-x}
dL\wedge \D(\vp)+d\left({}^{(bulk)}\!J\right)=0\,.
\end{equation}
The coordinate representation of this law can be written as 
\begin{equation}\label{J-cons-coord1-x}
L^{ij}{}_{,i}\vp_{,j} \d(\vp)+J^i{}_{,i}=0\,.
\end{equation}
This equation must be used, for example, when the surface absorbs charges from the medium.  
\subsection{Natural coordinates}
The surface current defined above is formally represented by 6 components of the antisymmetric tensor $L_{ij}$. However, not all of these components are independent. Indeed, the components of  $L$ which are tangential to the surface do not give a contribution to the surface current. Also  the ordinary description of the surface current  includes two components of a current vector $i_1,i_2$ tangential to the boundary surface  and one component of the charge density $\sigma$. 

First we note that due to the expression (\ref{delta-coord}), the 2-form $L$ is defined only up to an addition of an arbitrary 1-form multiplied by $d\vp$. Consequently, in the standard decomposition 
\begin{equation}\label{L-dec}
L=L_1+L_2\wedge d\vp\,
\end{equation}
the 1-form $L_2$ can be chosen to be equal to zero. Thus we can assume that the expansion of the 2-form $L$ does not involve the $d\vp$ factor. 

Define a coordinate system naturally adapted to the boundary surface
\begin{equation}\label{coord}
x^0=t\,,\quad x^1=y^1\,,\quad x^2=y^2\,,\quad x^3=\vp\,.
\end{equation}
In order to deal with the real boundary of a sample (not with an abstract surface) , we assume that the  hypersurface $\vp=0$ is transversal to the time axis. In other words,
\begin{equation}\label{coord1}
d\vp\wedge dt =\vp_{,\a}dx^\a\wedge dt\ne 0\,.
\end{equation}
Thus $\vp_{,\a}\ne 0$ at least for one value of $\a=1,2,3$.  

In this system, the coordinate decomposition of the 2-form $L$ is given by 
\begin{eqnarray}\label{L-dec1}
L&=&\frac 12 L_{ij}dx^i\wedge dx^j\nonumber\\&=&\frac 12 \left(L_{01}dt\wedge dy^1+L_{02}dt\wedge dy^2+
L_{12}dy^1\wedge dy^2\right)\,.
\end{eqnarray}
Since the current is localized on the surface, the components $L_{ij}$ are independent on the coordinate   $x^3=\vp$. 
Charge conservation law (\ref{J-cons}) takes now the form 
\begin{eqnarray}\label{L-cons}
dL\wedge \D(\vp)=\frac 12 \left(L_{01,2}-L_{02,1}+L_{12,0}\right)\d(\vp) dt\wedge dy^1\wedge dy^2\wedge d\vp=0\,.
\end{eqnarray}
It is useful to rewrite this equation  in  term of the dual tensor $L^{ij}=(1/2)\ep^{ijkl}L_{kl}$. We have the nonzero components $L_{01}=L^{23},L_{02}=-L^{13},L_{12}=L^{03}$, thus (\ref{L-cons}) yields
\begin{equation}\label{L-cons1x}
\left(L^{03}{}_{,0}+L^{13}{}_{,1}+L^{23}{}_{,2}\right)\d(\vp) dt\wedge dy^1\wedge dy^2\wedge d\vp=0\,.
\end{equation}
Denote the components as 
\begin{equation}\label{coord-x}
L^{03}=-\sigma\,,\quad L^{13}=i^1\,,\quad L^{23}=i^2\,.
\end{equation}
Thus, on the surface, the conservation law takes its ordinary form
\begin{equation}\label{L-cons2}
\frac {\partial i^1}{\partial y^1}+\frac {\partial i^2}{\partial y^2}=\frac {\partial \sigma}{\partial t}\,.
\end{equation}
\section{String electric current}
\subsection{String description and the corresponding Dirac delta-form}
In the rest position space,  a string is an one-dimensional object. Its motion is described in spacetime by a two-dimensional surface. It can be defined as an intersection of two hypersurfaces
 \begin{equation}\label{string1}
\left\{\begin{array}{ll}
 \vp(x^i)&=0\\
 \psi(x^i)&=0 \,.\end{array} \right.
\end{equation}
These hypersurfaces are assumed to be transversal to each  other 
 \begin{equation}\label{string2}
 d\vp\wedge d\psi\ne 0\,.
 \end{equation}
 Thus $\vp_{[,i}\psi_{,j]}\ne 0$ at least for one pair of indices $(i,j)$. Moreover,
 the hypersurfaces are assumed to be transversal to the time axis, i.e.,
 \begin{equation}\label{string3}
 d\vp\wedge dt\ne 0\,,\quad \mbox{and}\quad  d\psi\wedge dt\ne 0\,.
 \end{equation}
 Thus there is such pair of spatial indices $(\a,\b)$ that 
  \begin{equation}\label{string4}
 \vp_{,\a}\ne 0\,,\quad \mbox{and}\quad  \psi_{,\b}\ne 0\,.
 \end{equation}
 To define an electric charge localized on a string, we involve a 2-form
 \begin{equation}\label{string5}
 \D(\vp)\wedge\D(\psi)=\vp_{,i}\psi_{,j}\d(\vp)\d(\psi)dx^i\wedge dx^j\,,
 \end{equation}
 which is non-zero only on the intersection of the hypersurfaces $\vp$ and $\psi$. 
 \subsection{String electric current}
 Define a string electric current 3-form as 
 \begin{equation}\label{string6}
{}^{(str)}\!J=K\wedge\D(\vp)\wedge\D(\psi)\,,
 \end{equation}
 where $K$ is an arbitrary regular 1-form. (\ref{string6}) can be rewritten as 
 \begin{equation}\label{string7}
{}^{(str)}\!J=K\d(\vp)\d(\psi)\wedge d\vp\wedge d\psi\,,
 \end{equation}
 thus the current is tangential to the string
 \begin{equation}\label{string8}
{}^{(str)}\!J\wedge d\vp={}^{(str)}\!J\wedge d\psi=0\,.
 \end{equation}
 Moreover, ${}^{(str)}\!J$ has a support only on the solutions of the system (\ref{string1}), i.e., on the string itself. Since the factors $\D(\vp)$ and $\D(\psi)$ are dimensionless, the dimension of the 1-form $K$ is of charge. 
 
 For a coordinate expression, we write the 1-form $K$ as 
 \begin{equation}\label{string9}
K=K_idx^i\,.
 \end{equation}
 Thus, 
 \begin{equation}\label{string10}
{}^{(str)}\!J=K_i\vp_{,j}\psi_{,k}\d(\vp)\d(\psi)dx^i\wedge dx^j\wedge dx^k\,.
 \end{equation}
 Comparing with (\ref{curr}), we have 
 \begin{equation}\label{string11}
{}^{(str)}\!J_{ijk}=\frac 1{3!}K_{[i}\vp_{,j}\psi_{,k]}\d(\vp)\d(\psi)\,.
 \end{equation}
 A total charge containing in a 3-dimensional hypersurface $M_3$ is defined as 
 \begin{equation}\label{string12}
Q=\int_{M_3}{}^{(str)}\!J=\int_{M_3}K\wedge\D(\vp)\wedge\D(\psi)\,.
 \end{equation}
 Providing integration with respect to the factors $\vp$ and $\psi$ we remain with the expression
 \begin{equation}\label{string13}
Q=\int_{M_1}K\,,
 \end{equation}
 where $M_1$ is the intersection
 \begin{equation}\label{string14}
{M_1}=M_3\cap\{\vp=0\}\cap\{\psi=0\}\,.
 \end{equation}
 Thus the total electric charge of the string is exhibited by an integral of a 1-form. Already from this representation we can see that the form $K$  must be twisted otherwise the integral over a closed line will be identically equal to zero. Since the  dimension of $K$ is of charge, both sides of (\ref {string14}) have the proper physical dimension. 
 \subsection{String electric charge conservation}
 Using the representation (\ref{string6}) and the fact that the delta-form is closed, we derive the electric charge conservation law in the form 
 \begin{equation}\label{string15}
dK\wedge\D(\vp)\wedge\D(\psi)=0\,.
 \end{equation}
 In an arbitrary  coordinate chart, it takes the form
\begin{equation}\label{string16}
K_{[i,j}\vp_{,k}\psi_{,l]}\d(\vp)\d(\psi)=0\,.
\end{equation}

The integral representation of the conversational  law is obtained by the Stokes theorem. For a 3-dimensional closed surface $M_3$  bounded a 4-dimensional spacetime region,  
$M_3$ with a boundary $\partial M_4$,  it  takes the form
\begin{equation}\label{string17}
\int_{ M_3}K\wedge\D(\vp)\wedge\D(\psi)=0\,.
\end{equation}
Integrating with respect to the factors $\vp$ and $\psi$ we get 
\begin{equation}\label{string18}
\int_{ M_1}K=0\,,
\end{equation}
where the curve is defined as the intersection of the surfaces
\begin{equation}\label{string19}
M_1=M_3\cap\{\vp=0\}\cap\{\psi=0\}\,.
\end{equation}
Also in (\ref{string18}) the twisted nature of the 1-form $K$ is explicitly exhibited. 
\subsection{Natural coordinates for string}
In order to give an explicit meaning of the covariant formulas represented above, we introduce a system of coordinates adapted to a string:
 \begin{equation}\label{string20}
x^0=t\,,\quad x^1=\vp\,,\quad x^2=\psi\,,\quad x^3=y\,.
\end{equation}
The transversal conditions $dx^i\wedge dx^j\ne 0$ are assumed. The 1-form $K$ is defined in (\ref{string6}) only up to arbitrary additions of the terms proportional to $d\vp$ and $d\psi$. Thus we can restrict to the form $K$ which does not involve the differentials $dx^1$ and $dx^2$. Consequently the ordinary coordinate expression of  the 1-form $K$ is simplified as 
 \begin{equation}\label{string21}
K=K_0dt+K_3dy=jdt-\sigma dy\,.
\end{equation}
where we denoted the components $K_0=j$ for string electric current and $K_3=-\sigma$ for string electric charge. Take into account that both quantities are localized on the string, thus they independent on the variables $\vp$ and $\psi$. 

Thus we have the conservation law in the form 
\begin{eqnarray}\label{string22}
d\left({}^{(str)}\!J\right)&=&dK\wedge\D(\vp)\wedge\D(\psi)\nonumber\\
&=&\left(-\frac{\partial j}{\partial y}+\frac{\partial \sigma}{\partial t}\right)dt\wedge dy\wedge\D(\vp)\wedge\D(\psi)\,.
\end{eqnarray}
Consequently the conservation law takes the ordinary form
\begin{equation}\label{string23}
\frac{\partial \sigma}{\partial t}=\frac{\partial j}{\partial y}\,.
\end{equation}
\section{Point charges}
In order to complete our discussion of the singular low order charge densities, we consider the simplest but, probably, the most important case of  0-dimensional point charges. In spacetime, the point charge is described by a wordline which we will describe as the intersection of three hypersurfaces
 \begin{equation}\label{point1}
M_1=\{\vp(x^i)=0\}\cap\{\psi(x^i)=0\}\cap\{\tau(x^i)=0\}\,.
\end{equation}
Here three hypersurfaces are assumed to be pairwise  transversal with intersection on a 1-dimensional line. 
  
The electric charge 3-form is defined as 
 \begin{equation}\label{point2}
{}^{(point)}J=q \D(\vp)\wedge\D(\psi)\wedge\D(\tau)\,,
\end{equation}
where $q$ is a  pseudo-scalar $q$ which can be a function of a point. 
Explicitly, the current is rewritten as 
\begin{equation}\label{point3}
{}^{(point)}J=\Big(q \d(\vp)\d(\psi)\d(\tau)\Big)d\vp\wedge d(\psi)\wedge d(\tau)\,,
\end{equation}
In a coordinate charge, we have a representation  
  \begin{equation}\label{point5}
{}^{(point)}J=\Big(q \vp_{,i}\psi_{,j}\tau_{,k}\d(\vp)\d(\psi)\d(\tau)\Big)dx^i \wedge dx^j\wedge dx^k\,,
\end{equation}

The conservation law for the electric current takes the form
\begin{equation}\label{point5x}
dq \wedge\D(\vp)\wedge\D(\psi)\wedge\D(\tau)=0\,,
\end{equation}

To restate the ordinary representation we can chose the coordinates in such a way that 
\begin{eqnarray}\label{point6}
&&\vp_{,1}=1\,,\quad\vp_{,2}=0\,,\quad\vp_{,3}=0\,,\nonumber\\
&&\psi_{,1}=0\,,\quad\psi_{,2}=1\,,\quad\psi_{,3}=0\,,\nonumber\\
&&\tau_{,1}=0\,,\,\,\quad\tau_{,2}=0\,,\,\,\quad\tau_{,3}=1\,.
\end{eqnarray}
Thus we can write 
\begin{equation}\label{point7}
\vp=x-x_0(t)\,,\,\,\quad\psi=y-y_0(t)\,,\,\,\quad\tau=z-z_0(t)\,,
\end{equation}
where $x_0, y_0, z_0$ are arbitrary smooth functions of time (the constants of integration). The 4-vector $t,{\mathbf r}_0(t)$ with ${\mathbf r}_0(t)=\left(x_0(t),y_0(t),z_0(t)\right)$ represents now the wordline of the charge in spacetime. 
Consequently, 
\begin{equation}\label{point8}
\D(\vp)=\d(x-x_0(t))(dx-x'_0(t)dt)\,,
\end{equation}
and similarly for $\D(\psi)$ and $\D(\tau)$. 
Substituting into  (\ref{point5}) we have a representation  of  the form 
\begin{eqnarray}\label{point9}
{}^{(point)}J&=&\rho\d({\mathbf r}- {\mathbf r}_0(t))\Big( dx\wedge dy \wedge dz - x'_0(t)dt\wedge dy \wedge dz\nonumber\\&&+ y'_0(t)dt\wedge dx \wedge dz- z'_0(t)dt\wedge dx \wedge dy\Big)\,.
\end{eqnarray}
The dual components restate the ordinary representation of the 3D electric current, 
\begin{equation}\label{point10}
\rho=q\d({\mathbf r}- {\mathbf r}_0(t))\,,\qquad \mathbf j=q \mathbf v\d({\mathbf r}- {\mathbf r}_0(t))\,.
\end{equation}
Observe that these expressions are derived from the 3-form (\ref{point2}) and do not postulated ad hoc, compare \cite{LL2}. 
Moreover, using the relation $f(x)\d(x-x_0)=f(x_0)\d(x-x_0)$, we deduce that the scalar $q$ is independent on the spatial  coordinates. Now the charge conservation $dq=0$ yields that the charge must be independent also of time, i.e. to be a constant.

When the point charges are considered to be plunged into a charged medium, the charge conservation law picks up an additional  bulk term $dJ$
  \begin{equation}\label{point11}
\frac {\partial q}{\partial t}\, dt \wedge\D({\mathbf r}- {\mathbf r}_0(t))+dJ=0\,.
\end{equation}
This equation can be used to describe how  a pointwise charged aggregate is dissolved in a charged solution. 
\section{Conclusion}
One of the principle consequences of the diffeomorphism invariant  formulation of the electromagnetism is the fact that the electric current must be represented by a twisted 3-form. Such a form is a natural object for integration over 3-dimensional surfaces in space-time. In this paper we show that this notion is well defined also for the low dimensional electric charge densities. All the definitions are provided in a metric-free form. 

\section*{Acknowledgment}
  I would like to thank Friedrich Hehl  for his most fruitful
 comments.

\end{document}